\documentclass[useAMS,usenatbib]{mnras}

\usepackage[english]{babel}
\usepackage[utf8x]{inputenc}
\usepackage{graphicx}
\usepackage{color}
\usepackage{subfigure}
\usepackage{multirow}

\title{Faraday Rotation Measure Dependence on Galaxy Clusters Dynamics}
\author[Stasyszyn et al.]{F.A. Stasyszyn$^{1,2}$\thanks{E-mail: fstasyszyn@unc.edu.ar} \& M. de los Rios$^{1,2,3}$ \\
$^{1}$ Instituto de Astrof\'isica Te\'orica y Experimental (IATE), Laprida 854, C\'ordoba, Argentina\\
$^{2}$ Observatorio Astron\'omico de C\'ordoba, Universidad Nacional de C\'ordoba, Laprida 854, X5000BGR, C\'ordoba, Argentina.\\
$^{3}$ Consejo Nacional de Investigaciones Cient\'{\i}ficas y T\'ecnicas, Rivadavia 1917, C1033AAJ Buenos Aires, Argentina.\\
}

\begin{document}

\date{Accepted ???. Received ???; in original form ???}

\maketitle

\label{firstpage}

\begin{abstract}
We study the magnetic fields in galaxy clusters through Faraday rotation
measurements crossing systems in different dynamical states.
We confirm that magnetic fields are present in those systems and
analyze the difference between relaxed and unrelaxed samples with respect to
the dispersion between their inherent Faraday Rotation measurements. We found an
increase of this RM dispersion and a higher RM overlapping frequency for unrelaxed
clusters. This fact suggests that a large scale physical process is involved
in the nature of unrelaxed systems and possible depolarization effects are
present in the relaxed ones. We show that dynamically unrelaxed systems can
enhance magnetic fields to large coherence lengths. In
contrast, the results for relaxed systems suggests that small-scale dynamo can
be a dominant mechanism for sustaining magnetic fields, leading to intrinsic
depolarization. 
\end{abstract}


\section{Introduction}

The cosmological properties of magnetic fields (MF) and their influence in the
evolution of the Universe have become a relevant topic nowadays
\citep[i.e.][among others]{Kronberg2016, Beck2009ASTRA, Experiment2017,
2014ApJ...783L..20P, 2017MNRAS.469.3185P}.  Extensive studies have been carried out
\citep{Vacca2016} in order to estimate and derivate limits and properties of
cosmological magnetic fields.  In a previous investigation, \cite{Stasyszyn2010} studied
the relations between the cosmological large-scale structure (LSS) and the
Faraday rotation measurements using simulations. They found that the LSS produces a
signal that is below the current instrument's sensitivity thresholds and noise.  
Nevertheless, with the next generation of instruments, it should be
possible to apply statistical methods to detect signatures related with the large-scale
structure \citep{2015aska.confE..97V,2015aska.confE..95B,2016JApA...37...31K}.

Galaxy clusters are one of the astrophysical objects where MF have
been inferred without doubt \citep{carilli2002,feretti2012} and,
although we have obtained much information about their intrinsic physical
parameters \citep[i.e.][]{boehringer2010,kravtsovBorgani2012}, their role is still on
debate. For typical electron densities in clusters of galaxies, the expected
magnetic field should be able to generate a Faraday Rotation Measurement (RM)
signal that can provide information about the physics of the embedded plasma
\citep{stasyszyn2013,2017arXiv170703396M,Vazza2018}. For example, the RM magnitude and
coherence length are related and give us information about the turbulence energy
of the system, which helps us to infer the possible explanations about a dynamo
effect preserving those fields.

One of the first studies to address a statistical study on the intra-cluster
medium (ICM) is the one published by \cite{clarke2001}, where they probed the magnetic field strength
over 16 galaxy clusters and found an excess of RM towards the center of the
clusters.  By simply modeling the electron column density from X-ray observations,
they deduced an average magnetic field strength of  $<|B|> = 5 - 10~ (l/10 {\rm
kpc})^{1/2}~ h^{1/2}_{75} \mu G$.  From this research to nowadays, both, the number
of extragalactic RM observations and the information about clusters have been
improved.  \cite{boehringer2016} studied the correlation of the observed RM
with cluster properties using $1722$ X-ray luminous galaxy clusters from {\rm
CLASSIX}.  These clusters were identified down to a nominal lower flux-limit
of $1.8~10^{-12} {\rm erg s^{-1} cm^{-2}}$ in the $0.1~-~2.4~{\rm keV}$ energy
band.  In their analysis, they searched for the closer RMs to the center of each
cluster. By modeling the electron column density from the clusters and assigning
this RM measurement to it, they found an RM dispersion that characterizes the
sample.  They confirmed previous results and found that the RM	 dispersion
increases with the column density, inferring a value of $<|B|> = 2 - 6~ (l/10
{\rm kpc})^{1/2}\mu$G. 

\citet{Bonafede2011} made a similar analysis, but measuring lower values of
polarization fraction towards the center of galaxy clusters, which can be
explained by MF models with central values of $\mu$G.  In their research, they
were able to subdivide the samples taking into account the presence of radio
halos, cool core and high and low temperatures.  They did not find any
difference in the depolarization for the temperature and radio halos cases, but
they did for the cool core and non-cool core samples.

In this paper, we use the statistics of RMs in the lines of sight of galaxy
clusters to infer properties related to the magnetic field in clusters.
Specifically, we use dynamical information that allows us to divide the sample
into relaxed and unrelaxed clusters, and study the relations between their
magnetic fields and their dynamical status.

The paper is organized as follows. In Section \ref{sec:data}, we present the
observational samples and the data that we use in our analysis. In Sections
\ref{sec:analysis} and \ref{sec:stat}, we present the statistical analysis and
the main results of the present work.  Finally, in Section \ref{sec:discussion} and
Section \ref{sec:conclusions}, we discuss some important aspects of our results
and we present a few concluding remarks.


\section{Observational samples} \label{sec:data}

In this study we statistically infer the relation between the magnetic field
properties and the dynamical status of different galaxy clusters samples.  To
this end, we correlate the positions of the clusters in the samples with the RMs
around them and study the collision chance between a given Galaxy cluster and
the RM distribution.

\subsection{Galaxy cluster samples}

\cite{wen} studied galaxy clusters subsamples from the SDSS/DR8, and classified
them into relaxed and unrelaxed using photometric information.  The relaxed
galaxy clusters refer to the ones that are not interacting and that should have
a smooth symmetrical mass and light distribution, while the unrelaxed clusters
are the ones that have important substructures that may come from strong
interaction with other clusters of similar characteristics.  It is well-known
that strong interactions between clusters introduce very important
modifications in all the components of the systems (galaxies, intra-cluster
medium, dark matter distribution, etc.). This produces different signatures
in all the electromagnetic spectrum (presence of radio halos, asymmetric X-ray
distribution, etc.).  Using this fact, \cite{wen} calibrated their method with a
sample of galaxy clusters using multiband information and a very well-known
dynamical state.

They estimated 3 parameters that quantify different properties of the light
distribution of clusters, namely \footnote{For the exact definitions of these
parameters, please read \cite{wen}.}:

\begin{itemize}
\item $\alpha$: This parameter quantifies the asymmetry of the galaxy
distribution.  
\item $\beta$: The smoothed optical map of relaxed cluster has a
steep surface brightness profile in all directions, while on the other hand,
the presence of substructures imprints a 'ridge' in a certain direction in the
smoothed map.  This parameter quantifies the difference in the light profile in
the ridge direction compared to others.  
\item $\delta$: As relaxed clusters have a very similar light profile in all
directions, their optical map can be fitted by two-dimensional elliptical King
model, while, on the other hand, clusters with a lot of substructures would
deviate more from this model. The normalized deviation $\delta$ quantifies the
deviation of the optical map of the cluster from this two-dimensional King
model.  
\end{itemize}

Finally, by using these parameters and the clusters with well-known dynamical
status, they defined the relaxation parameter $\Gamma$ as
\begin{equation}
\Gamma = \beta - 1.90\alpha+3.58\delta + 0.10
\end{equation}
and found that relaxed clusters have $\Gamma > 0$ and unrelaxed clusters have $\Gamma < 0$.

\begin{figure}
\centering
\includegraphics[width=0.22\textwidth]{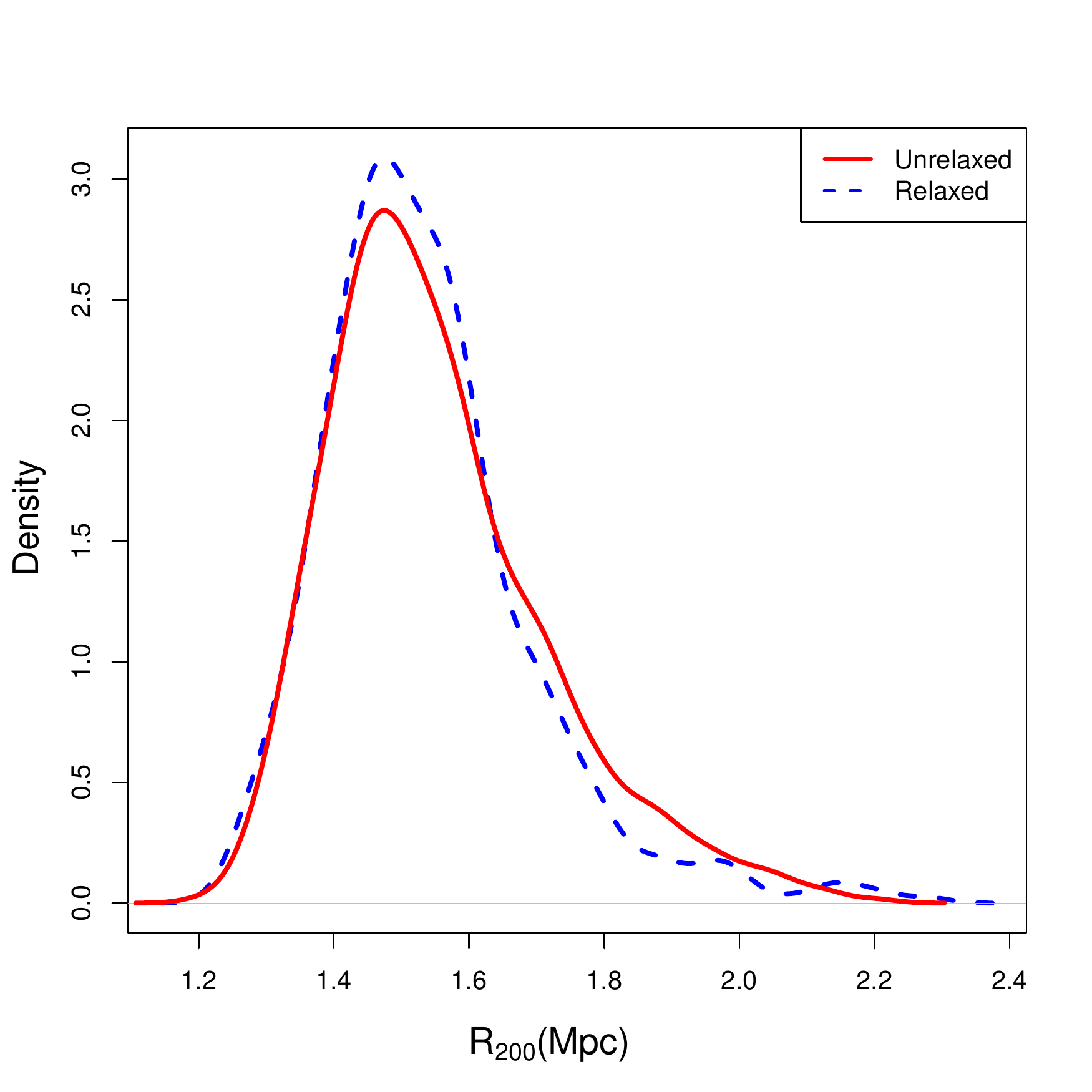}
\includegraphics[width=0.22\textwidth]{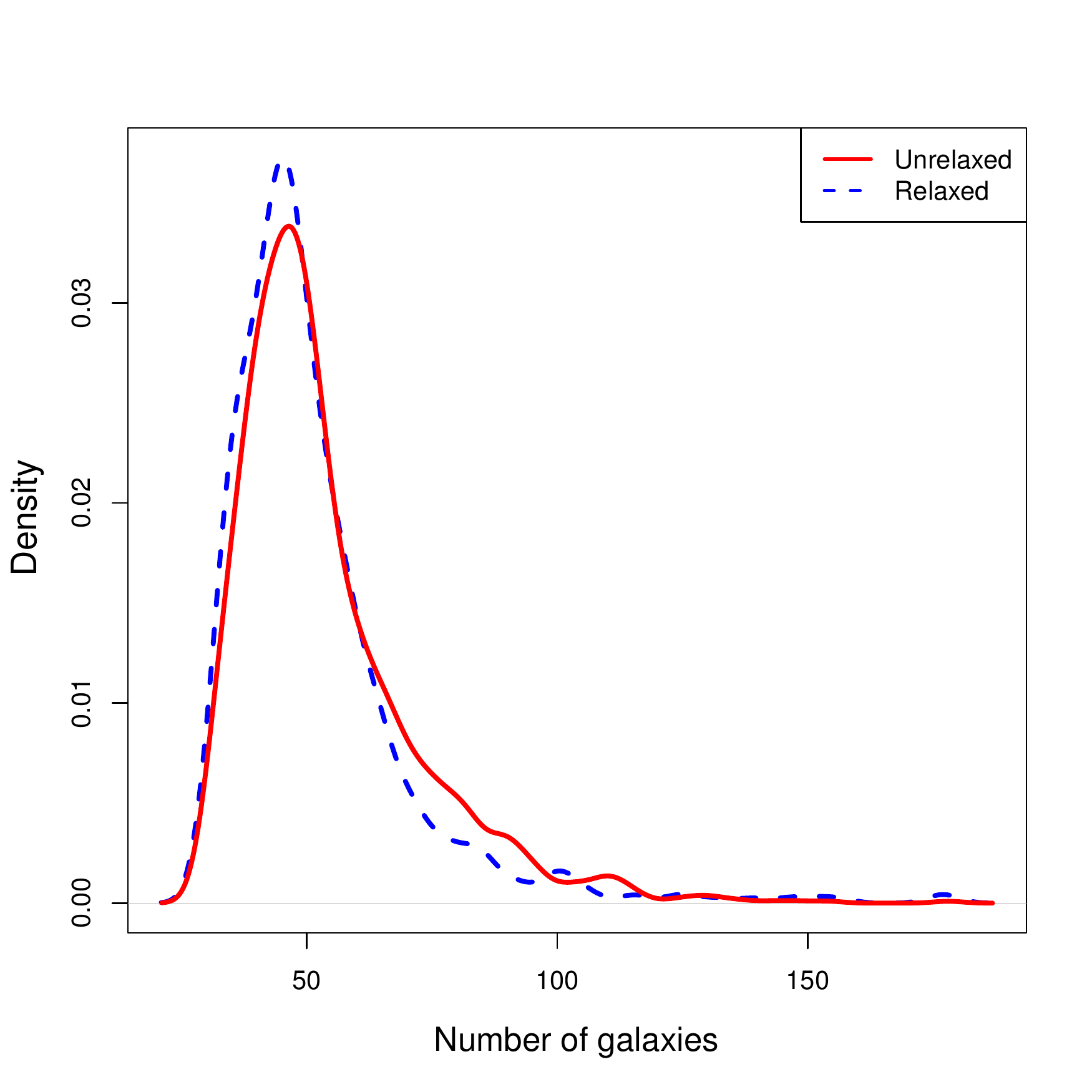}
\includegraphics[width=0.22\textwidth]{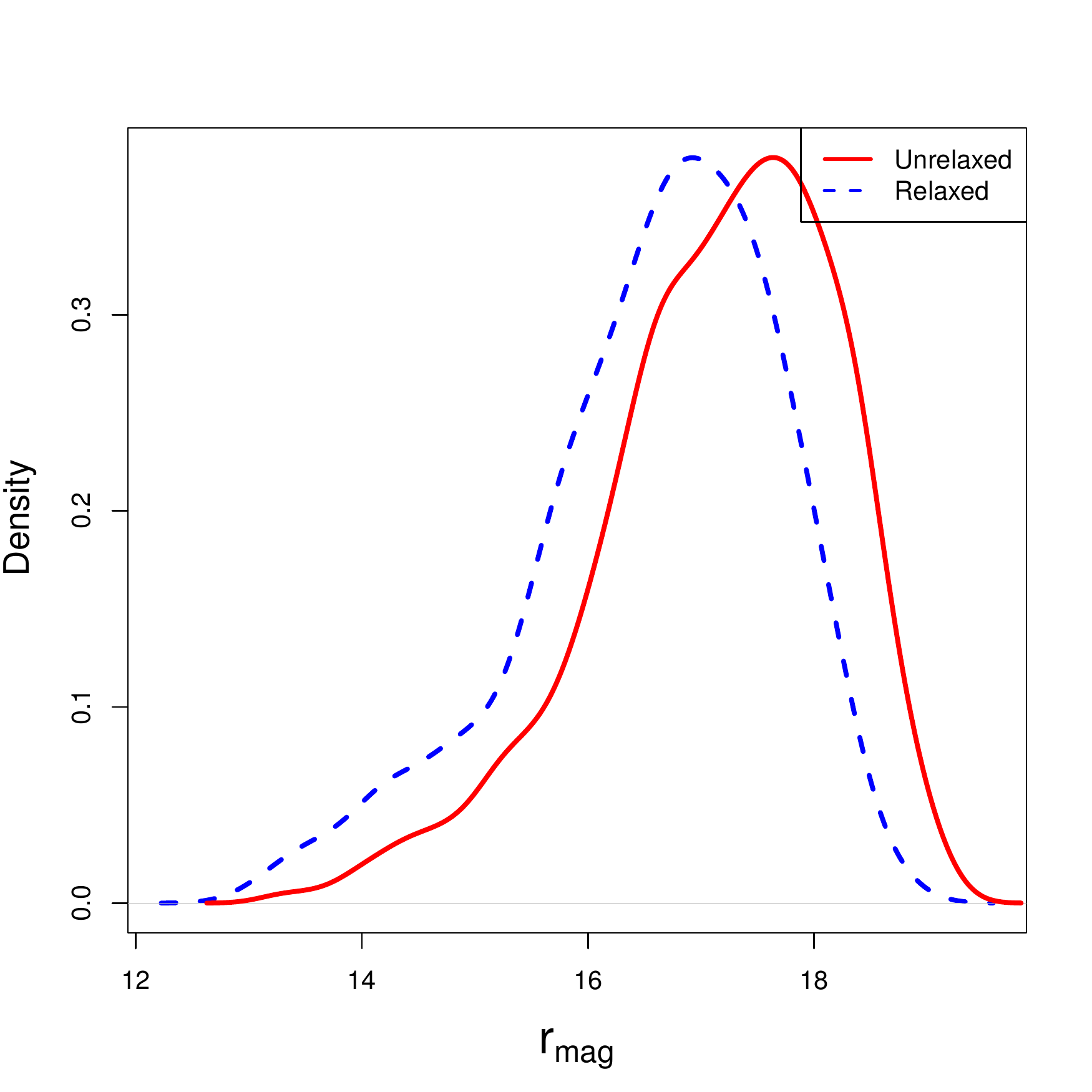}
\includegraphics[width=0.22\textwidth]{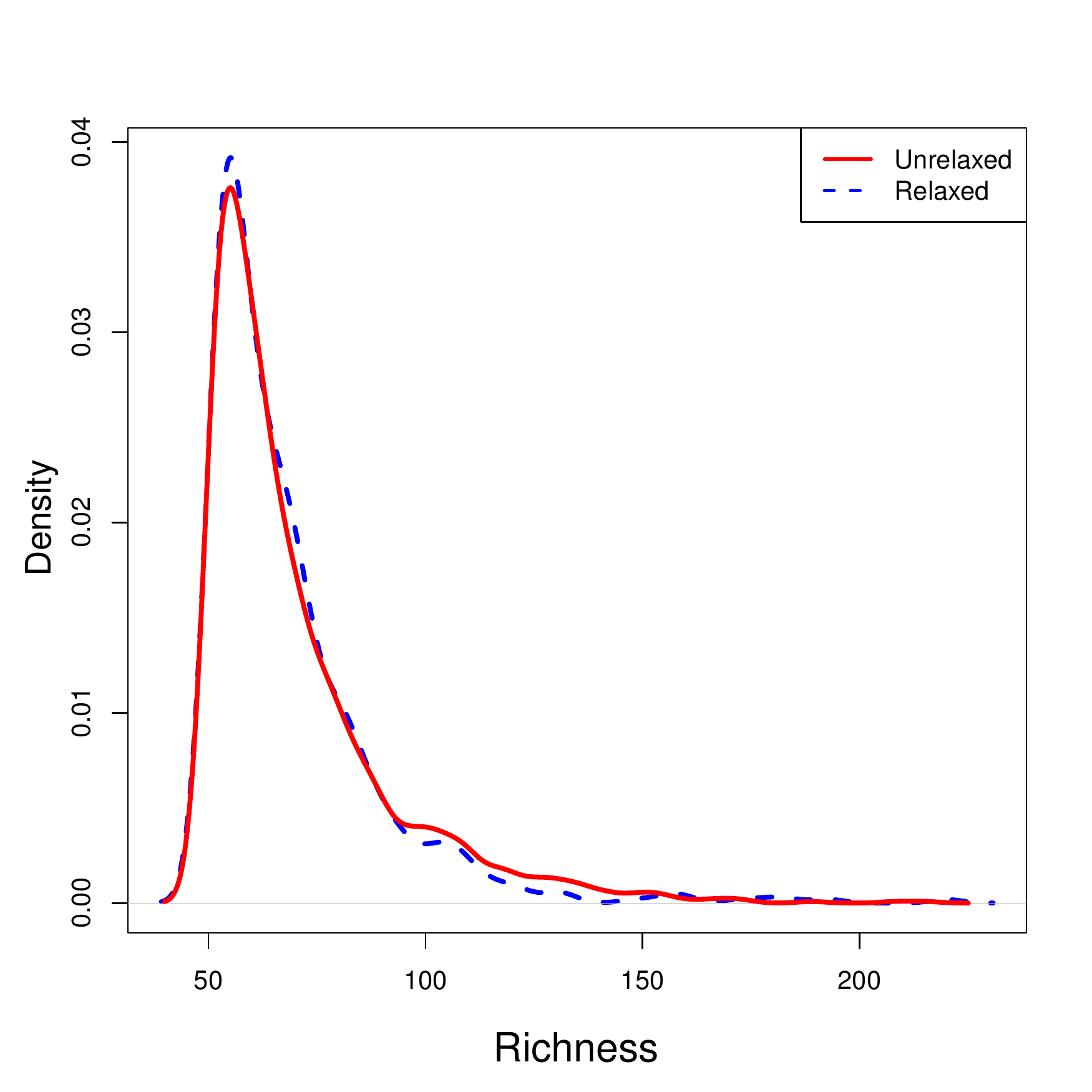}
\includegraphics[width=0.22\textwidth]{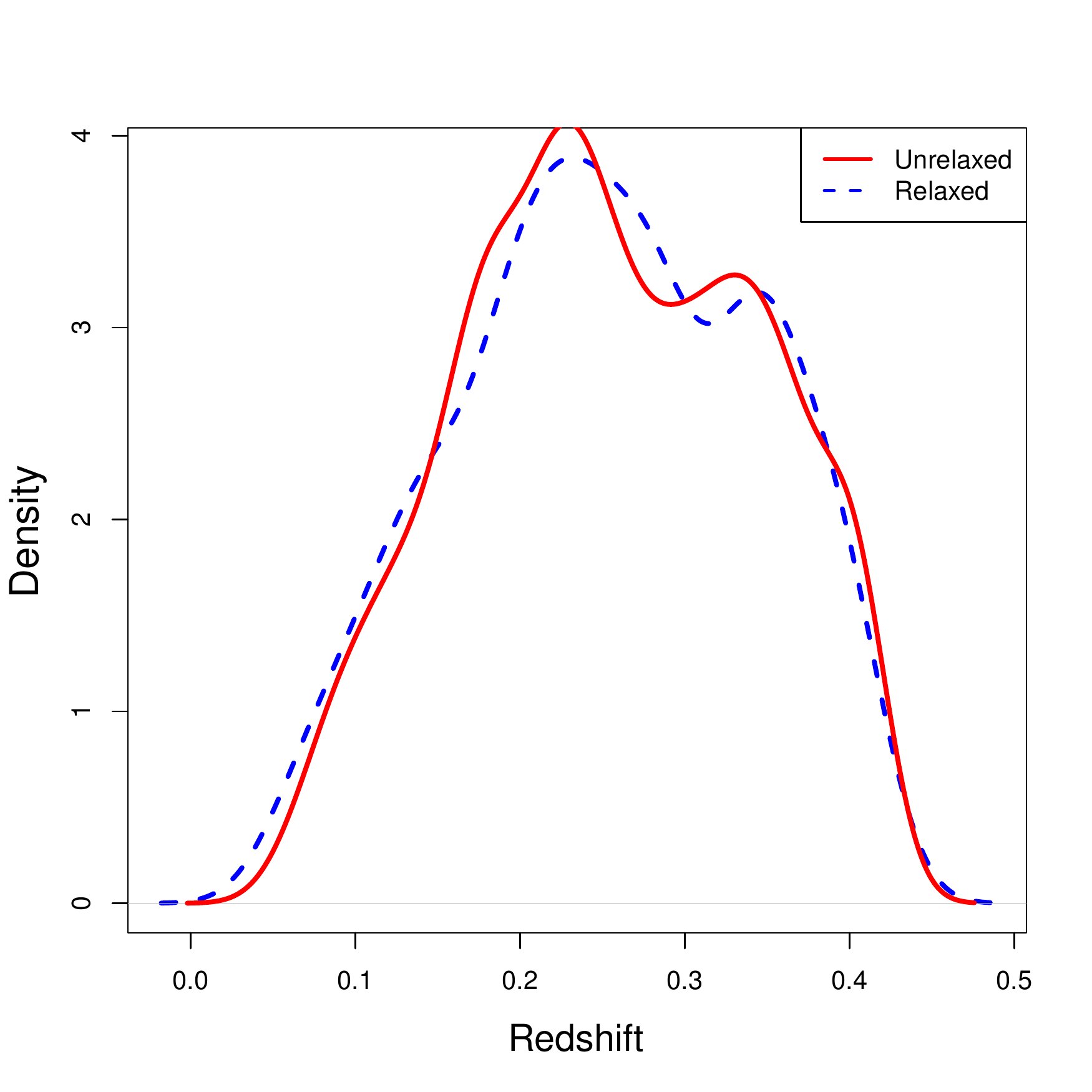}
\includegraphics[width=0.22\textwidth]{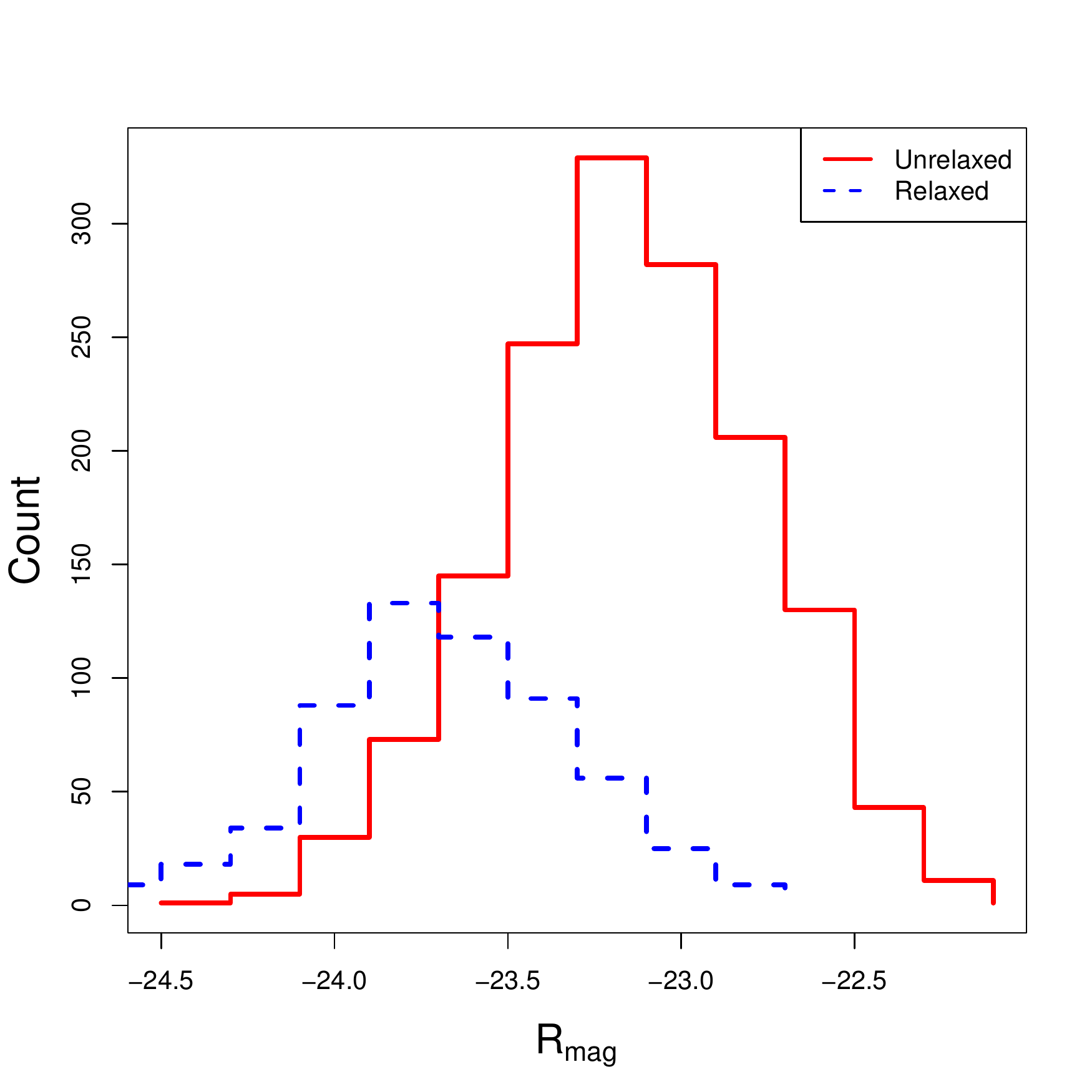}
\caption[Distribution properties form the samples]
{Comparison of the properties of relaxed and unrelaxed galaxy clusters samples. 
\textit{Top left}: Cluster radius ($r_{200}$) distribution. \textit{Top right}: Number of Galaxies distribution. \textit{Center left}: Apparent R magnitude distribution. \textit{Center Right}: Cluster Richness distribution. 
\textit{Bottom left}: Redshift distribution. \textit{Bottom right}: Absolute magnitude histogram.
}
\label{fig:distributions}
\end{figure}

In the final sample, we found $589$ in the relaxed state ($\Gamma > 0$) and
$1503$ unrelaxed clusters ($\Gamma < 0$).  In figure \ref{fig:distributions}, we
show the distributions of the main properties of the two samples.  We found
that all the distributions are very similar with the exception of the $R_{mag}$,
where the relaxed clusters are slightly more brilliant.  This can be understood
because, even if they have similar masses, the {\it relaxed} clusters already
have formed stars. Meanwhile, in the {\it unrelaxed} or {\it merging} clusters, the
star formation is just set up by the merger. Excluding the $R_{mag}$
distribution, the two samples
only differ in their dynamical state.

In Table 1 of \citet{wen}, they described the galaxy cluster dynamical
information that they used as a known learning sample for their neural network.
Additionally to the relaxed or unrelaxed state, they included information of whether 
they are known to have cool core, radio halo/relic or if they have confirmed 
merging signatures. It can be shown that the merging cases and the ones with radio 
halo/relic are always present in the case classified as unrelaxed. On the other 
hand, the cool core signature is present in both states, being a $25\%$ of the 
relaxed cases. 

We also performed a cross-correlation between the relaxed an unrelaxed cluster
with AGN catalogs of different bands. Specifically, we used the catalog of
confirmed AGN of \cite{Veron2010}, the WISE catalog of candidates to AGN
\citep{wise}, the $\gamma$-Ray AGN catalog of \cite{abdo} and the radio AGN
catalog of \cite{horiuchi}. In order to study the correlations between galaxy
clusters and AGNs, we counted how many clusters have at least one AGN inside a
radius of $30'$.  In all the catalogs, we found that both clusters samples have
the same ratios of systems with AGNs at all radii.  We summarize these results
in Table \ref{agn}.

\begin{table}
\caption{
Cross-correlation between galaxy clusters and AGN catalogs of different bands.
}
\label{agn}
\begin{tabular}{lrr}
\hline
\noalign{\smallskip}
Catalog & Unrelaxed & Relaxed \\
\noalign{\smallskip}
\hline
\noalign{\smallskip}
\cite{Veron2010} &  $72\%$ & $74\%$ \\
\hline
\noalign{\smallskip}
\cite{wise} &  $97\%$ & $96\%$ \\
\hline
\noalign{\smallskip}
\cite{abdo} &  $1.9\%$ & $1.8\%$ \\
\hline
\noalign{\smallskip}
\cite{horiuchi}  & $0.8\%$ & $0.3\%$ \\
\hline
\end{tabular}
\end{table}

\subsection{Faraday rotation measurements observations}

The other ingredient for our study is the RM information.  We used the catalog
of \cite{taylor2009} that contains Faraday rotations measures of $37,543$
polarized radio sources from the NRAO VLA Sky Survey (NVSS). Although this
catalog is obtained using only two frequencies (which can lead to ambiguities
for high values of RM), it has a large survey area that allows us to perform
a better correlation of the distribution of galaxy clusters from 
catalogs.

It is worth to note that there are only two extreme RM in the area covered by
our clusters sample. The ambiguities in the RM determination do not statistically
contribute and, hence, have no influence on our following analysis.
We took into account the foreground contribution of our Galaxy by
subtracting the average RM value in $6$ degrees around each system, as done by
\cite{boehringer2016}, \cite{Stasyszyn2010} and others.

In addition to \cite{taylor2009}, we also considered \cite{xu2014} work and
performed an analysis using their RM catalog. Although it is smaller, it 
has the advantage of having identified the extragalactic sources; therefore,
having the redshift of the RM source. This allows us to analyze only the sources
that are behind (or in front of) each cluster and so, to have more accurate inference
of the Faraday depth. However, it does not guarantee  that we drop sources that
are just not identified with a cosmological counterpart. 

\section{Analysis} \label{sec:analysis}

\begin{figure}
\centering
\includegraphics[width=0.4\textwidth]{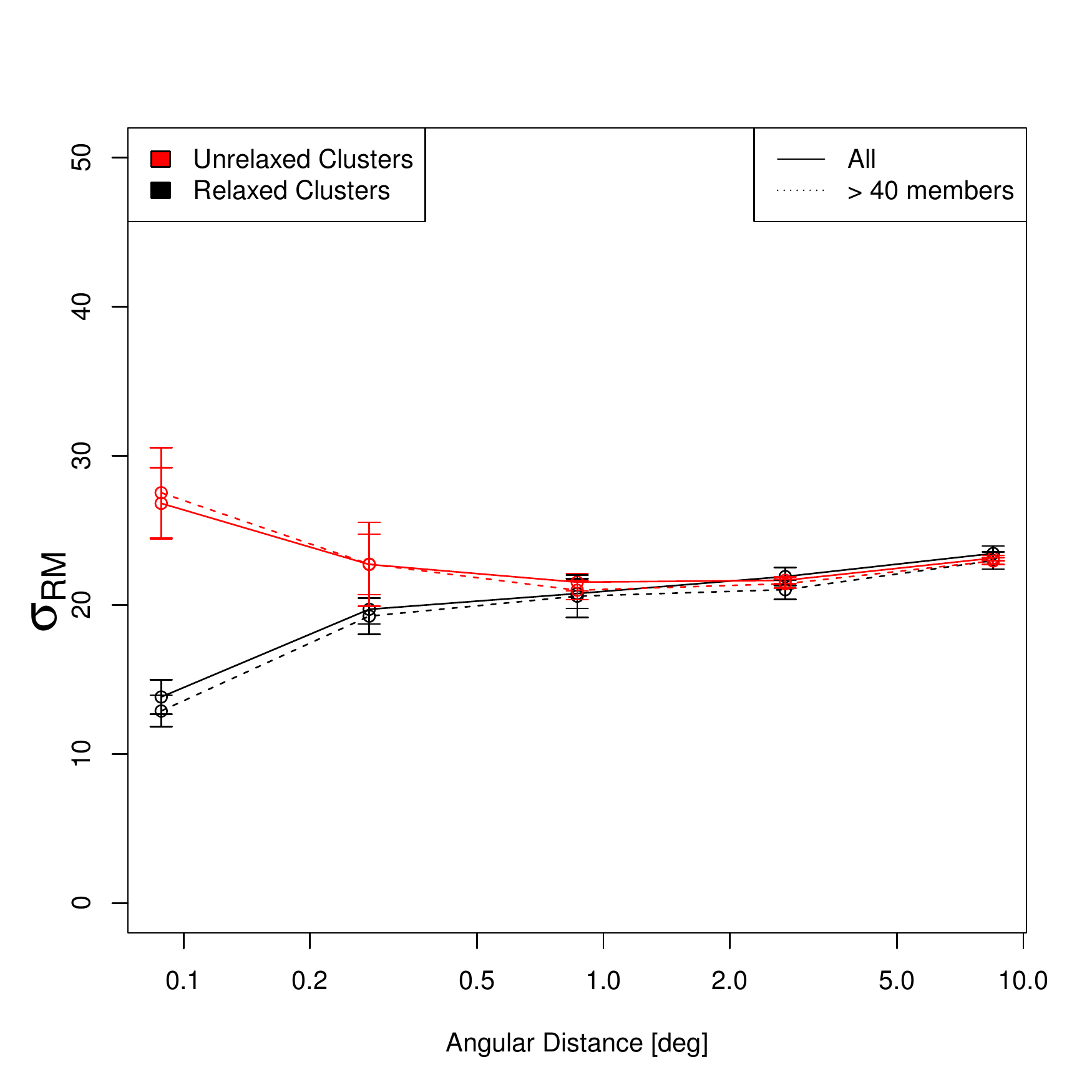}
\caption[angular sigma(RMs)]
{Faraday rotation standard deviation as a function of angular distance of the galaxy cluster samples.
We show the unrelaxed clusters in red and the relaxed ones in black.
The solid lines correspond to the full samples and 
the dashed lines correspond to the ones restricted to have more than 40 members.
}
\label{fig:angular}
\end{figure}

We started our analysis measuring the standard deviation of the RMs
($\sigma_{RM}$) of the relaxed and the unrelaxed clusters in $5$ radial bins
around the direction of each cluster.  In figure \ref{fig:angular}, we show the
standard deviation of the RM as a function of the angular distance for the full
samples (solid line).  In order to check if this difference is dominated by
smaller groups of galaxies, we performed the same calculation, but only with
clusters with more than $40$ galaxies, and displayed this results in dashed
lines.  It can be seen that in both samples the unrelaxed clusters
statistically have a larger RM dispersion and that the difference is larger
when we impose more than $40$ members. This implies that the effect is mostly
driven by the bigger galaxy clusters.  To estimate the errors of our
measurements, we used a bootstrapping technique that consist in estimating the
$\sigma_{RM}$ removing one of the galaxy clusters each time, and then we
calculate the dispersion of these measurements.

In order to avoid biases caused by the difference in the luminosity, and taking
advantage of the fact that the unrelaxed clusters over count in a factor of 3 the relaxed
ones (shown in figure \ref{fig:distributions}), we decided to randomly
subsample the unrelaxed clusters to follow the $R_{mag}$ distribution and
numbers density of the relaxed ones.  As done previously, we estimated the
standard deviation of the RM and found that the unrelaxed systems have
larger RM dispersions. It is worth noting that this is the same trend found in
the analysis of the full sample, indicating that the difference between
the relaxed and unrelaxed clusters is not driven by the number of galaxy
clusters neither by the $R_{mag}$ distribution.

\begin{figure}
\centering
\includegraphics[width=0.4\textwidth]{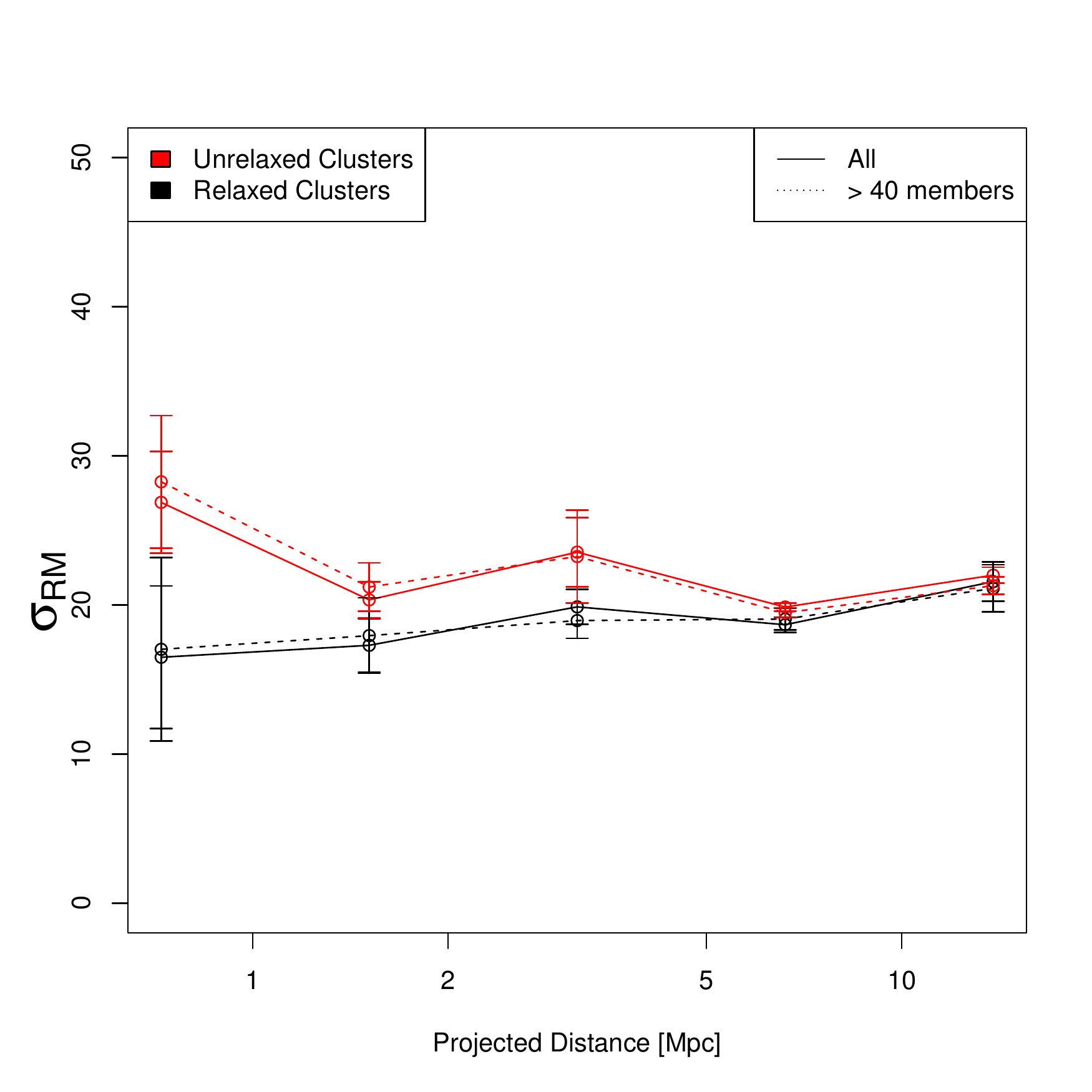}
\caption[projected sigma(RMs)]
{Faraday rotation standard deviation as a function of projected distance to the galaxy cluster samples. Same colors and lines reference as in figure \ref{fig:angular}.
}
\label{fig:projected}
\end{figure}

In figure \ref{fig:projected}, we show the same calculations but projecting to
the physical distance of the central cluster. Again, the solid line corresponds
to the full samples, while the dashed lines correspond to clusters with more
than 40 galaxy members.  We found the same trend as in the previous plot, the
unrelaxed clusters have statistically a larger RM dispersion than the relaxed
ones.

\begin{figure}
\centering
\includegraphics[width=0.4\textwidth]{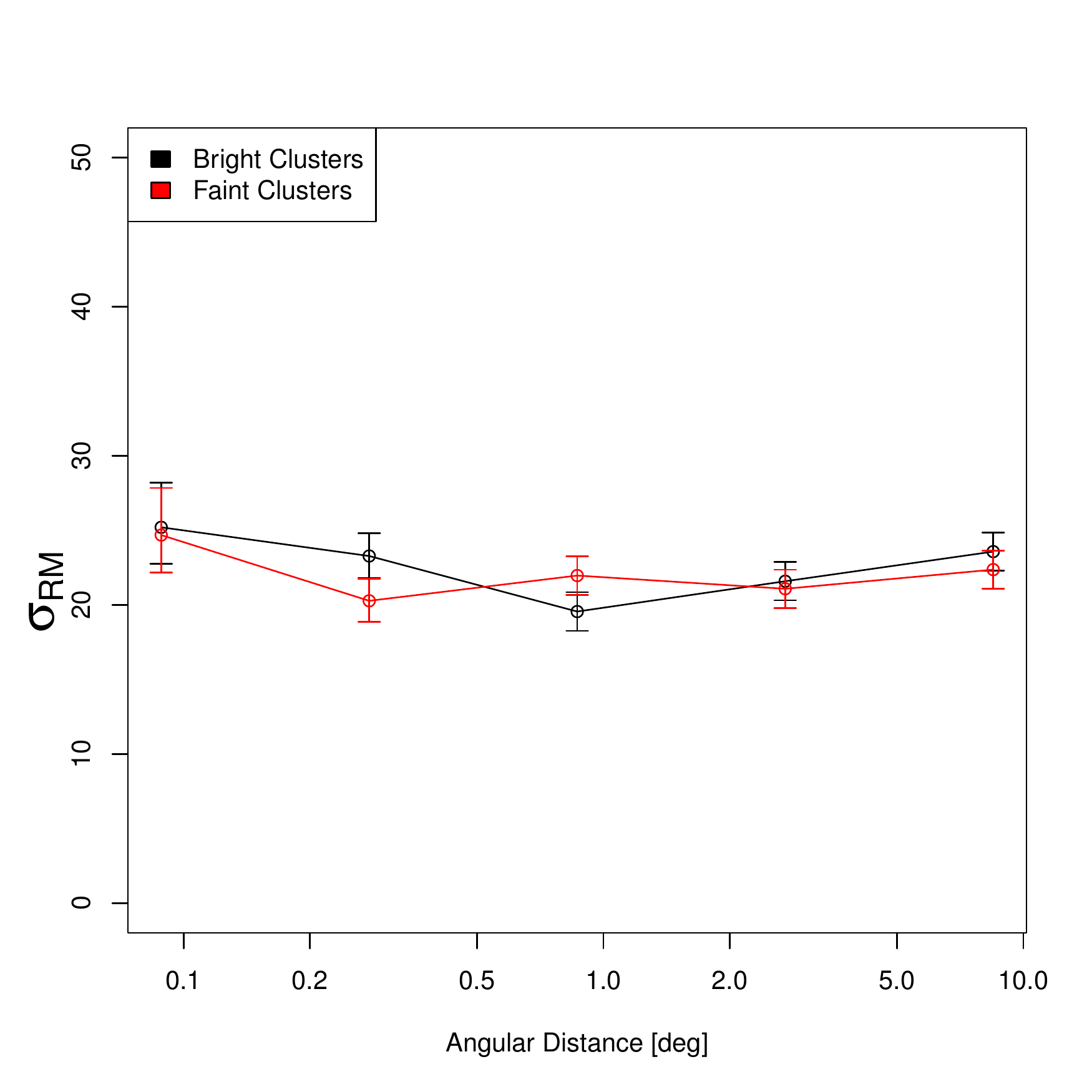}
\caption[Angular Bright Faint]
{Faraday rotation standard deviation as a function of angular distance to the galaxy cluster samples. 
The bright clusters are shown in the black lines and the fainter ones are shown in the red lines.
We do not observe any significant difference.
}
\label{fig:bri}
\end{figure}

As it can be seen in all the plots, the difference between samples rises when
we select clusters with more than $40$ members. This result can be understood because the small
clusters do not really contribute to this effect, due to the fact that the small
mass of those objects and the small intervening volume do not generate a
relevant Faraday rotation effect.  To continue the analysis of the samples, we
divided the galaxy clusters into bright and faint (making the cut in
$-23.5~R_{mag}$) without using the dynamical state information and performed
the same calculation of the RM dispersion in both sub-samples.  We displayed
these results in figure \ref{fig:bri}, where we observe that there are no
differences between these two samples, confirming that the difference displayed
in previous plots is due to their dynamical state.

\begin{table}
\caption{
Standard deviation of the observed rotation measure (in $rad/m^2$) in different rings around the galaxy clusters line of sight direction.
}
\label{tab1}
\[
\begin{array}{lrrr}
\hline
\noalign{\smallskip}
{\rm Angular} &   <0.5 Deg & 0.5-1.5 Deg & 1.5-3 Deg  \\
\noalign{\smallskip}
\hline
\noalign{\smallskip}
{\it Relaxed} & {20.0} \pm 1.0 & 20.8 \pm 1.0 & 21.9 \pm 0.6 \\
{\it Unrelaxed} & {23.2} \pm 1.7 & 21.7 \pm 0.6 & 21.5 \pm 0.2 \\
\noalign{\smallskip}
\hline
\noalign{\smallskip}
{\rm Projected} & < 1.0 {\rm Mpc} & 1.0 - 5.0  {\rm Mpc}  & 5.0 - 10 {\rm Mpc}  \\
\noalign{\smallskip}
\hline
\noalign{\smallskip}
{\it Relaxed} & 18.7 \pm 3.4 & 19.7 \pm 1.0 & 18.8 \pm 0.5  \\
{\it Unrelaxed} & 25.3 \pm 2.1 & 22.9 \pm 1.6 & 19.7 \pm 0.3  \\
\noalign{\smallskip}
\hline
\noalign{\smallskip}
\end{array}
\]
\end{table}

\begin{figure}
\centering
\includegraphics[width=0.4\textwidth]{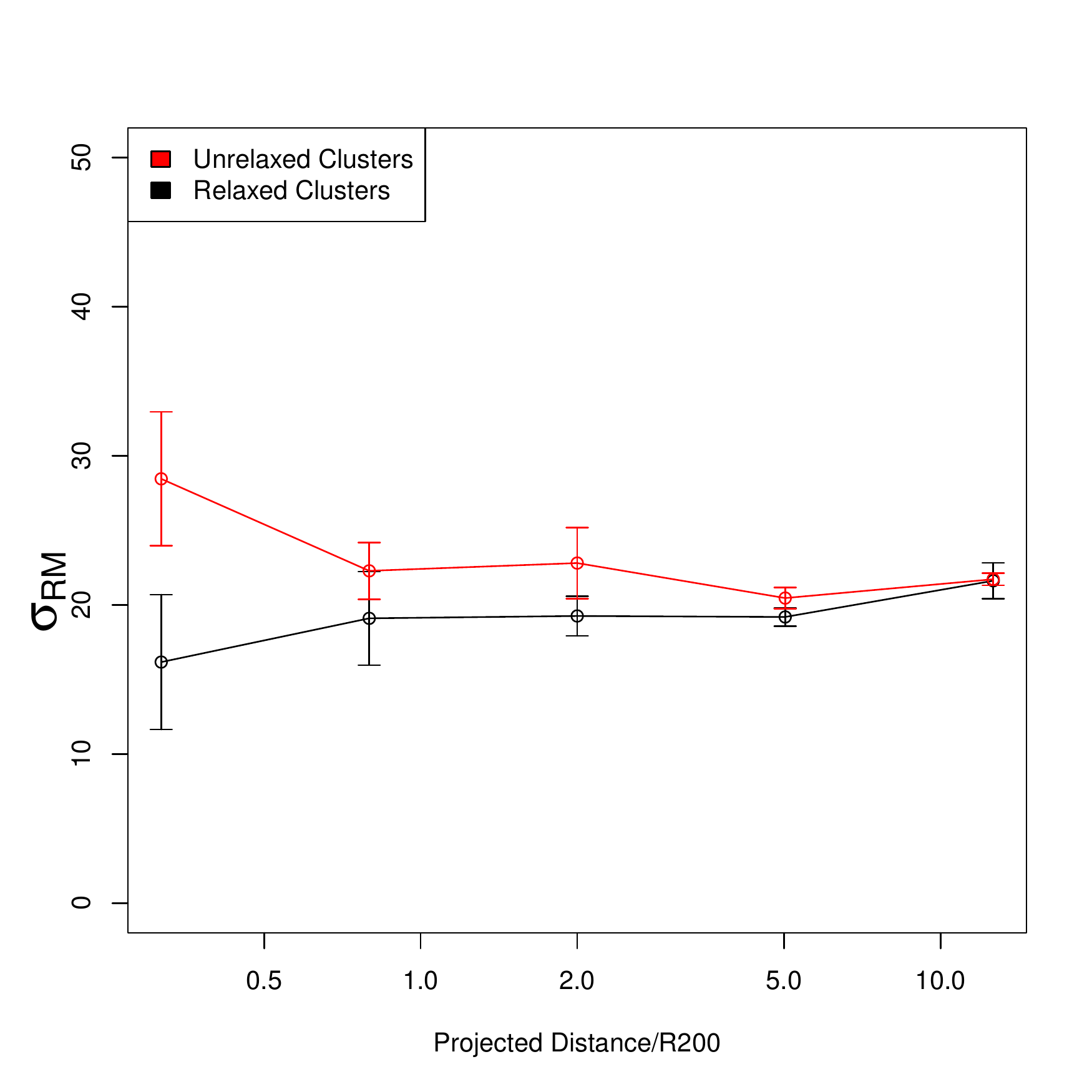}
\caption[projeted Sigma normalized R200]
{Faraday rotation standard deviation as a function of projected distance normalized to the $R_{200}$ to the galaxy cluster samples calculated by \cite{wen}. 
Same colors and line reference as figure \ref{fig:angular}.
}
\label{fig:r200}
\end{figure}

In figure \ref{fig:r200}, we renormalized the projected distance to the $R_{200}$
estimated by \cite{wen}.  We can observe that the difference holds up to twice
the physical radius, showing that it is inherent to the cluster
themselves and their surrounding medium.

\begin{figure}
\centering
\includegraphics[width=0.4\textwidth]{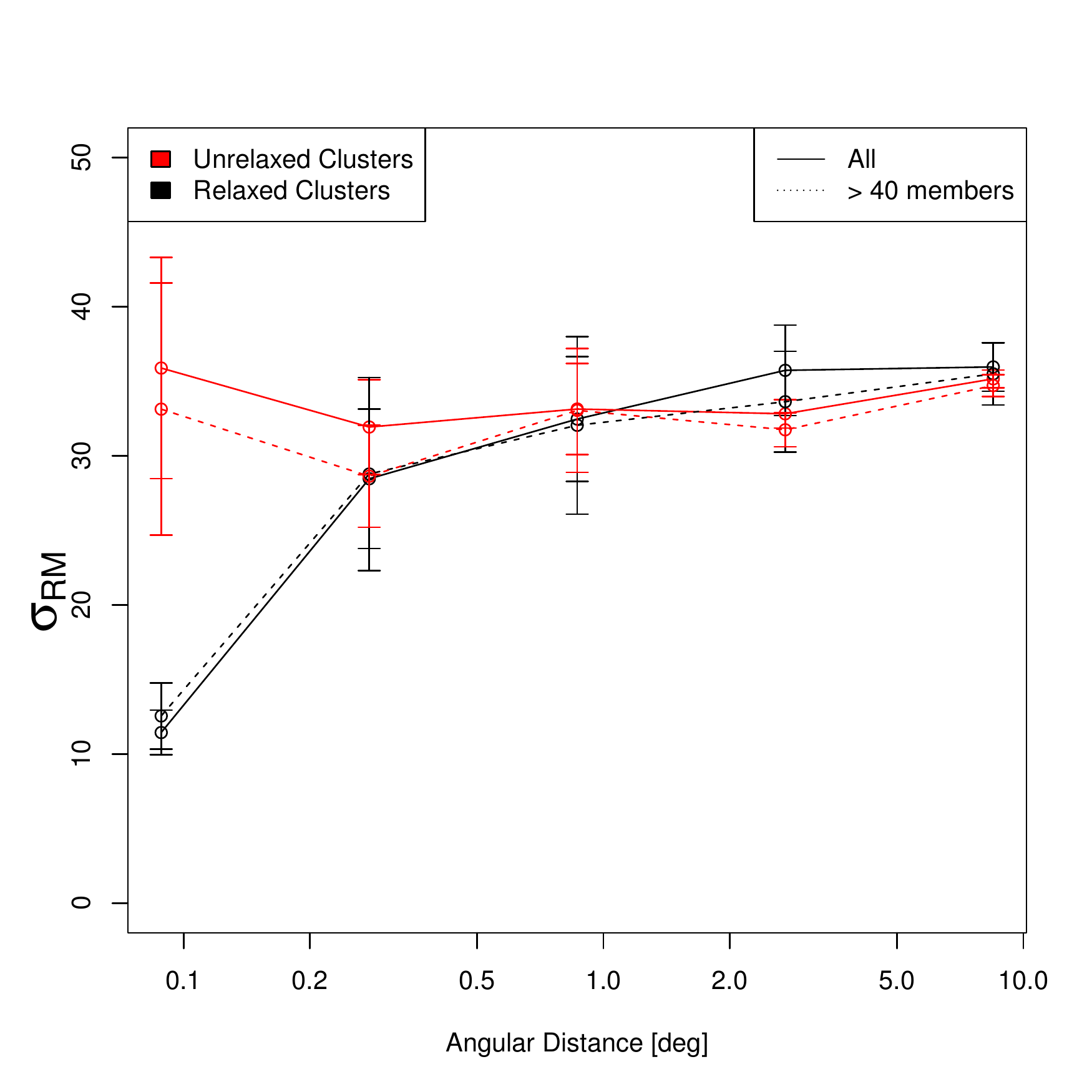}
\caption[Angular sigma(RMs) by Han]
{Faraday rotation standard deviation as a function of angular distance to the galaxy cluster samples using the \cite{xu2014} RM catalog.
Here we are taking into account only the samples that are behind the galaxy cluster samples.
Same colors and line reference as figure \ref{fig:angular}.
}
\label{fig:AngHan}
\end{figure}

As the RM is an integrated effect along the line of sight that includes all the
cosmic structures between the polarized source and the cluster, to not know the
distance of the polarized sources is a strong source of uncertainties in our
method.  \cite{xu2014} compiled a catalog of $\sim 3600$ confirmed
extragalactic RM sources, for which they also measured a redshift.  Therefore,
we performed our test discarding all the RM sources that are in front of the
galaxy clusters.  The result is shown in figure \ref{fig:AngHan}, where we show
the standard deviation of the RM as a function of the projected distance to
clusters centers.  It can be seen that they follow the same trend as previous
analysis. However, as the number of line of sights sources is drastically
reduced (it is just $6\%$ of the \cite{taylor2009} sample), the error bars are
bigger.

\section{Statistical density distributions} \label{sec:stat}

\begin{figure}
\centering
\includegraphics[width=0.4\textwidth]{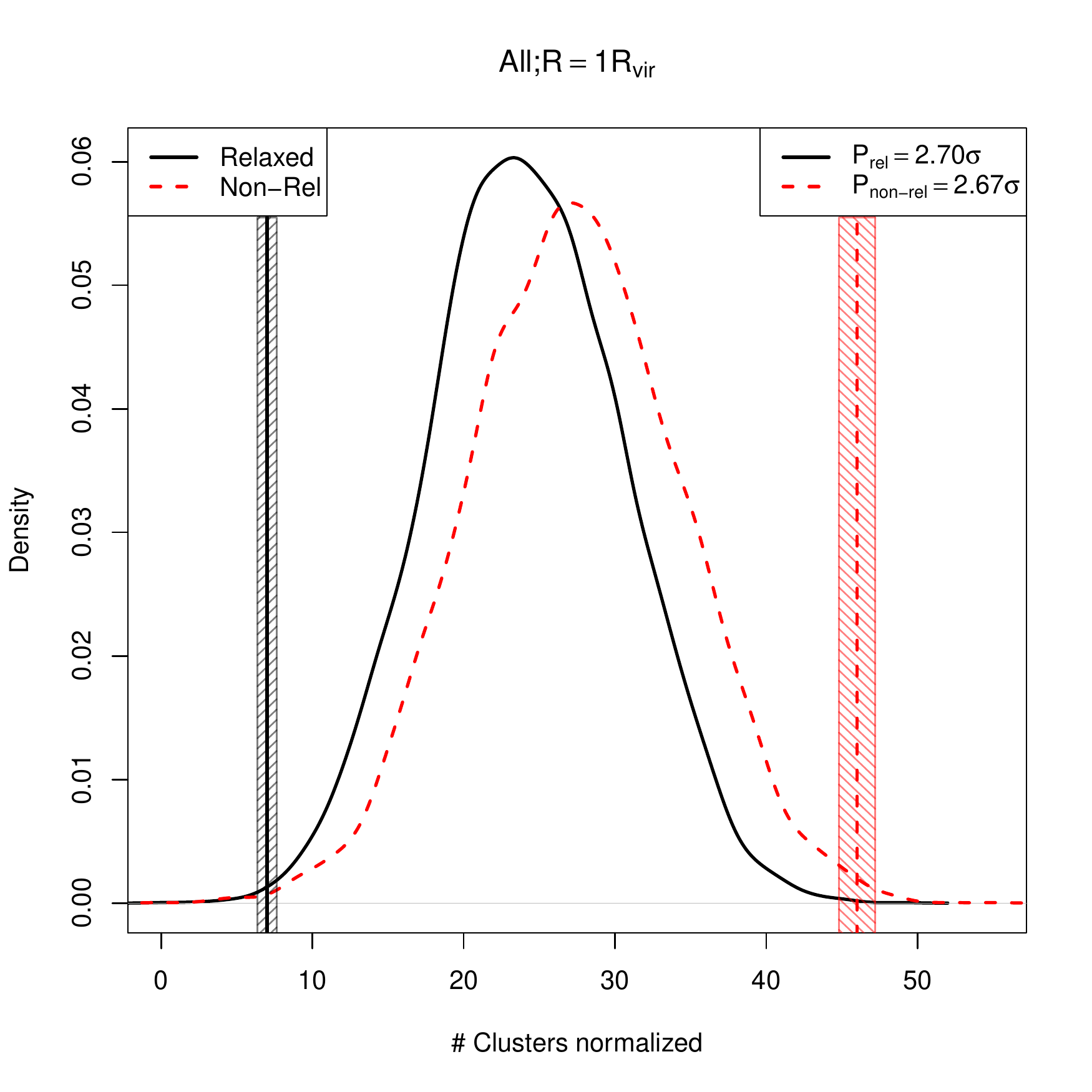}
\caption[all density distribution]
{Probability distribution of having at least one line of sight RM within one virial radius of each galaxy clusters in $1000$ randomly constructed 
catalog realizations.  
In black and red are plotted the relaxed and unrelaxed cluster samples respectively while in vertical lines are plotted the real catalog measurements.
For comparison between the samples we subtracted the minimum of crossing RMs to each one.
}
\label{fig:density}
\end{figure}

\begin{figure}
\centering
\includegraphics[width=0.4\textwidth]{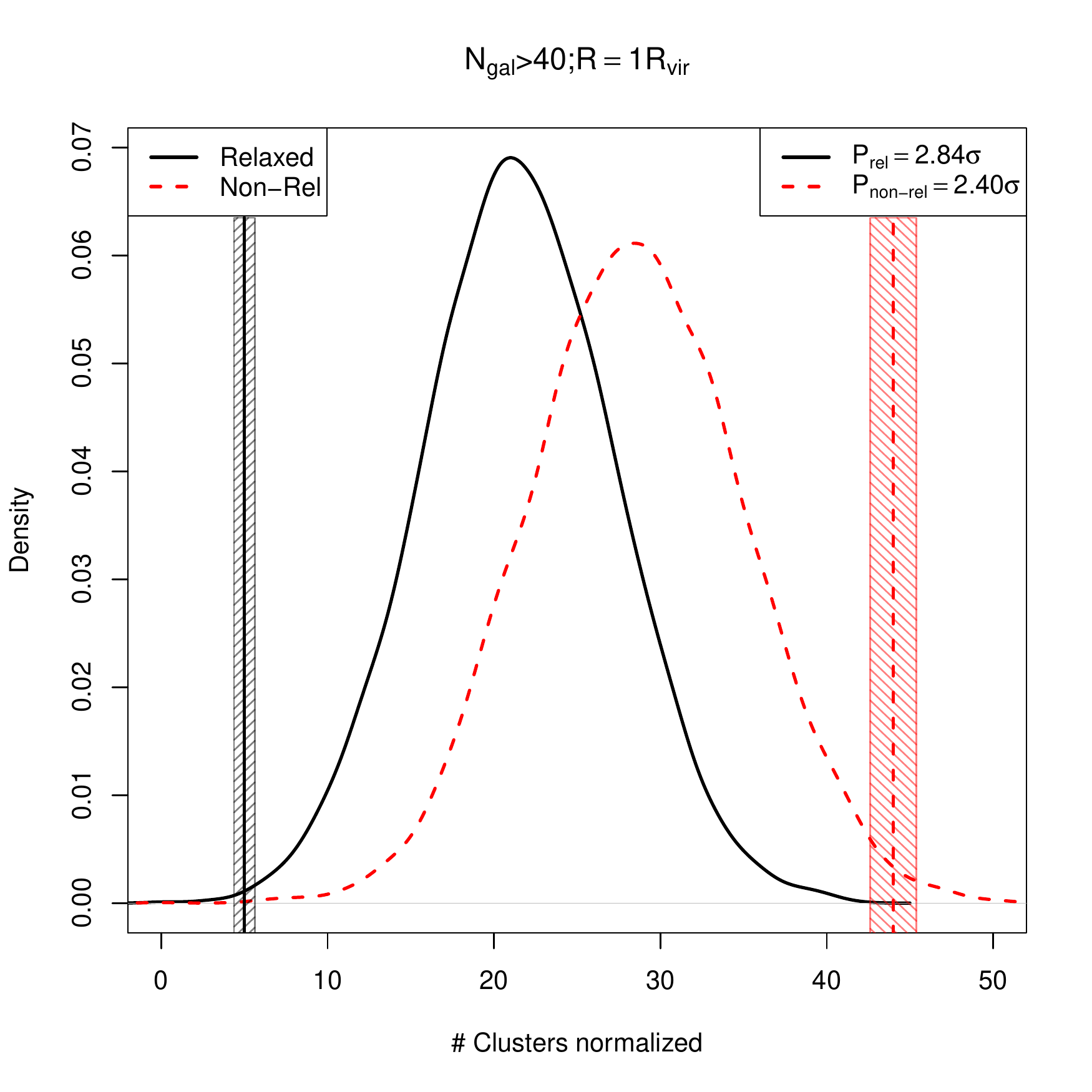}
\includegraphics[width=0.4\textwidth]{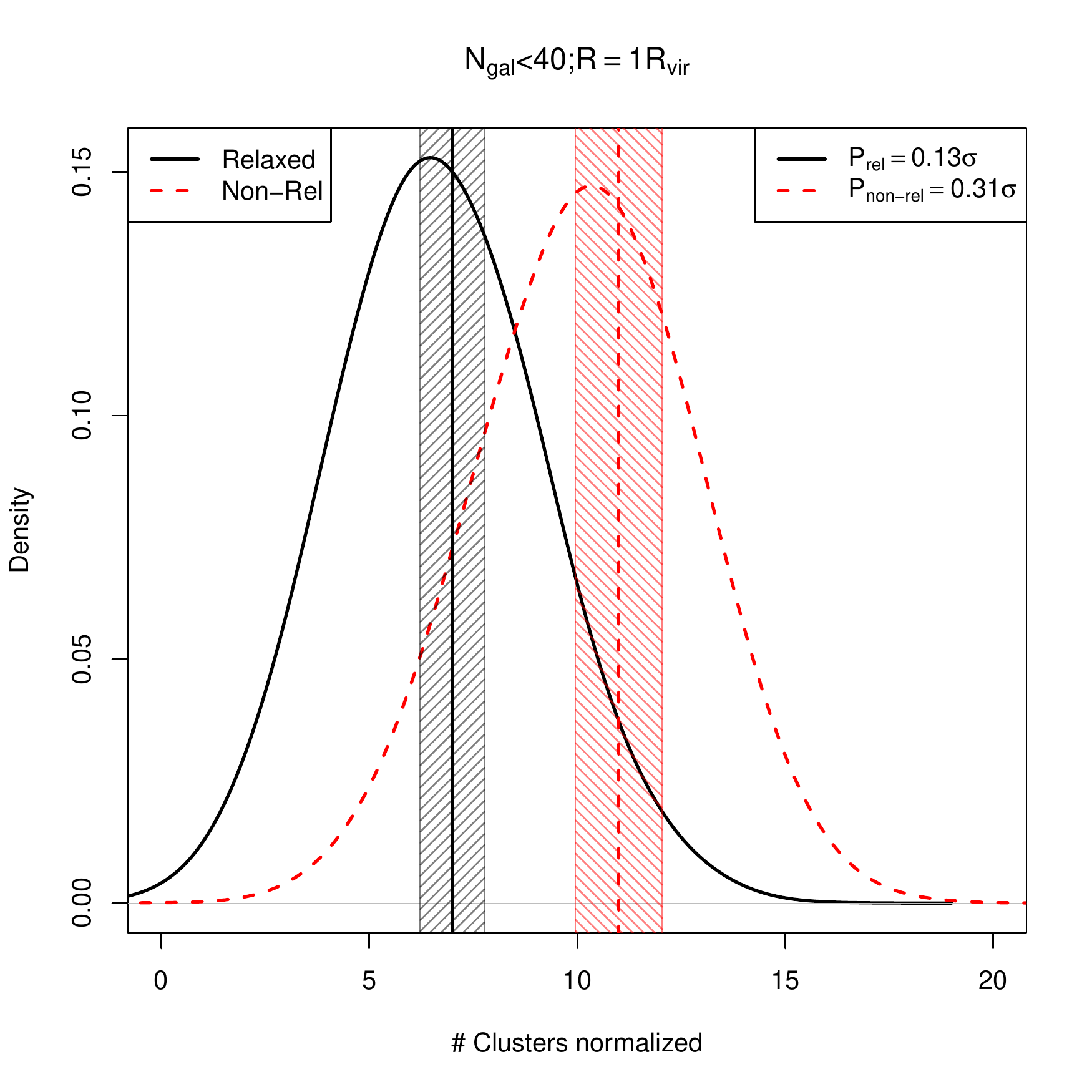}
\caption[Density ditribution M40]
{Probability distribution of having at least one line of sight RM within one virial radius of each galaxy clusters with more 
(less) than $40$ members in $1000$ randomly constructed catalog realizations.
The relaxed and unrelaxed cluster samples are plotted in black and red, respectively, while the real catalog measurements are plotted in vertical lines.
In order to compare between the samples, we subtracted the minimum of crossing RMs to each one. Same colors and line reference as figure \ref{fig:density}.
}
\label{fig:density2}
\end{figure}

We studied the probability of line of sight overlapping galaxy clusters with the
rotation measure positions. To achieve this, we generated $1000$ catalogs randomly sorting
the angular positions of the galaxy clusters and analyzed the probability of
having at least one RM within one virial
radius of each cluster. The results are shown in the figure \ref{fig:density}.
There, we compare the probability distribution of the random realizations with
the results obtained in the relaxed (black solid line) and unrelaxed (red dashed line) samples.
It can be seen that the relaxed and unrelaxed observations are almost
$3~\sigma$ off the random realizations and in opposite trends.  
To compare the distributions (that have different numbers of galaxy
clusters), we subtracted to each of the dynamical samples the minimum RM line of
sight numbers. We also estimated the error of the observational measurements
using the bootstrapping method as in the previous section.  This demonstrates that
relaxed clusters have a smaller overlapping chance compared with a random
distribution and the unrelaxed cases have a higher chance of having at least
one RM crossing them.  If we now only consider clusters with more (less) than
$40$ members, we obtain the plots of figure \ref{fig:density2}. We show that
galaxy clusters with more than $40$ members are also significantly different
than the random samples, whereas those with less than $40$ members do not show
a big difference compared with the random distributions.

This is consistent with previous sections in which we show that the RM
dispersion increases and decreases for the same cases.  We show that the
frequency of having at least one RM crossing the clusters is higher for the
unrelaxed clusters than for the random sample, and is smaller for the relaxed
ones.

\section{Discussion} \label{sec:discussion}

\citet{carilli2002} reviewed the possible origins of magnetic field. They
described the variations that exist in field strengths and topologies,
especially when comparing dynamically relaxed clusters to those that have
recently undergone a merger. They pointed out that, in all cases, the MF
has a significant effect on energy transport in the intracluster medium.
\citet{Subramanian2006} extensively discussed analytical models for the origin of
turbulence and magnetic fields in galaxy clusters, having good agreement in
their estimations compared with observations. Recently, in \citet{Vazza2018}, they were
able to resolve small scale dynamos acting on the intercluster medium.  All
these studies, between others, help us to understand to some extent the
complexity of galaxy clusters and the astrophysics involved.  For the
statistical nature of the present work we decided to take a rougher approach in
these discussions, leaving the detailed modeling for further work.

As in other studies \citep[i.e.]{boehringer2016}, we can assume that the observed
RM originates from the integration in the line of sight with different magnetic
field orientations and ICM cells.  The $\sigma_{RM}$ is diluted by the
characteristic length of the ICM cell that passes through the observed line of
sight by $\Lambda = (L/l)^{1/2} $ , where $L$ is the length of the ICM column
electron density and $l$ is the characteristic length of the ICM cell with
coherent magnetic field.  From this, we are able to derive a relation between
the $\sigma_{RM}$, the electron density $N_e$ and the magnetic field
$B_\parallel$ as : 

\begin{equation}
 \frac{B_{||} }{1 \mu {\rm G}} = 3.801 \times 10^{18} \frac{\sigma_{RM}}{{\rm rad~m}^{-2}} ~ \left(\frac{N_e}{{\rm cm}^{-2}}\right)^{-1}~\Lambda  
\label{eqn:BfromSigma}
\end{equation}

As a difference from \cite{boehringer2016}, we do not have a good estimator for
the $N_e$ for each sample.  Using as characteristic values $N_e = 10^{21}
cm^{-2}$, $L \sim 1~{\rm Mpc}$ and $l \sim 1~{\rm Kpc} $, one can infer values
of $3.0 \pm 0.25 \mu {\rm G}$ for the unrelaxed and $2.2 \pm 0.40 \mu {\rm G}$
for the relaxed systems.  If one also assumes $l = 10 ~{\rm Kpc}$ for the
unrelaxed clusters and $l = 25 ~{\rm Kpc}$ for the relaxed with the same cell
size of $L=1~{\rm Mpc}$, the outcome values for the magnetic fields are
$B_\parallel = 0.96\pm 0.08~ \mu {\rm G} $ and $B_\parallel = 0.45 \pm 0.08 \mu
{\rm G} $, respectively.  We used the same $L$ value because we did not find a
significant difference in the distribution properties (figure
\ref{fig:distributions}). However, we expect that the turbulence will be higher
for the unrelaxed sample, therefore implying a smaller $l$.  Note that if
we consider different values for $N_e$, we expect a larger difference
between the samples, noting that it is proportional with this quantity. We also
know that we expect the relaxed clusters or the ones with cooling cores having
higher densities; therefore, there will be a decrease in the inferred MF.
However, we empathize again that the way the cluster samples were selected does not
guarantee that cooling core clusters are in one side of the set samples.
Comparing with \cite{boehringer2016}, we have a difference of almost one order of
magnitude in the $\sigma_{RM}$ values. This is related with the fact that the
samples that we used are not the same (neither the RM sources nor the Galaxy
clusters) and have different statistical properties compared with the ones used
in their work.

One interesting aspect of our results is that if it is true that the RM
dispersion is smaller in the relaxed cases, this can also be due to the
depolarization effect from the intrinsic sources \citep{Bonafede2011}. This means
that we are not able to observe the Faraday effect, but not that the MF
is not present. It can also explain the results from section
\ref{sec:stat},  where we found that it is less frequent to have an RM crossing the relaxed
samples compared with random samples. We can justify this just by the fact
that those clusters have just depolarized the rotation measures and we cannot
observe them.  However, as a difference with \cite{Bonafede2011}, there are differences in the
way we divided the samples (relaxed and unrelaxed). 
From the learning dataset, the unrelaxed are the only ones to have
radio halos or relics, and about $50\%$ of the relaxed samples have cool
core measured, while just an $8\%$ of the unrelaxed have been identified as cool cores.  

\begin{figure}
\centering
\includegraphics[width=0.4\textwidth]{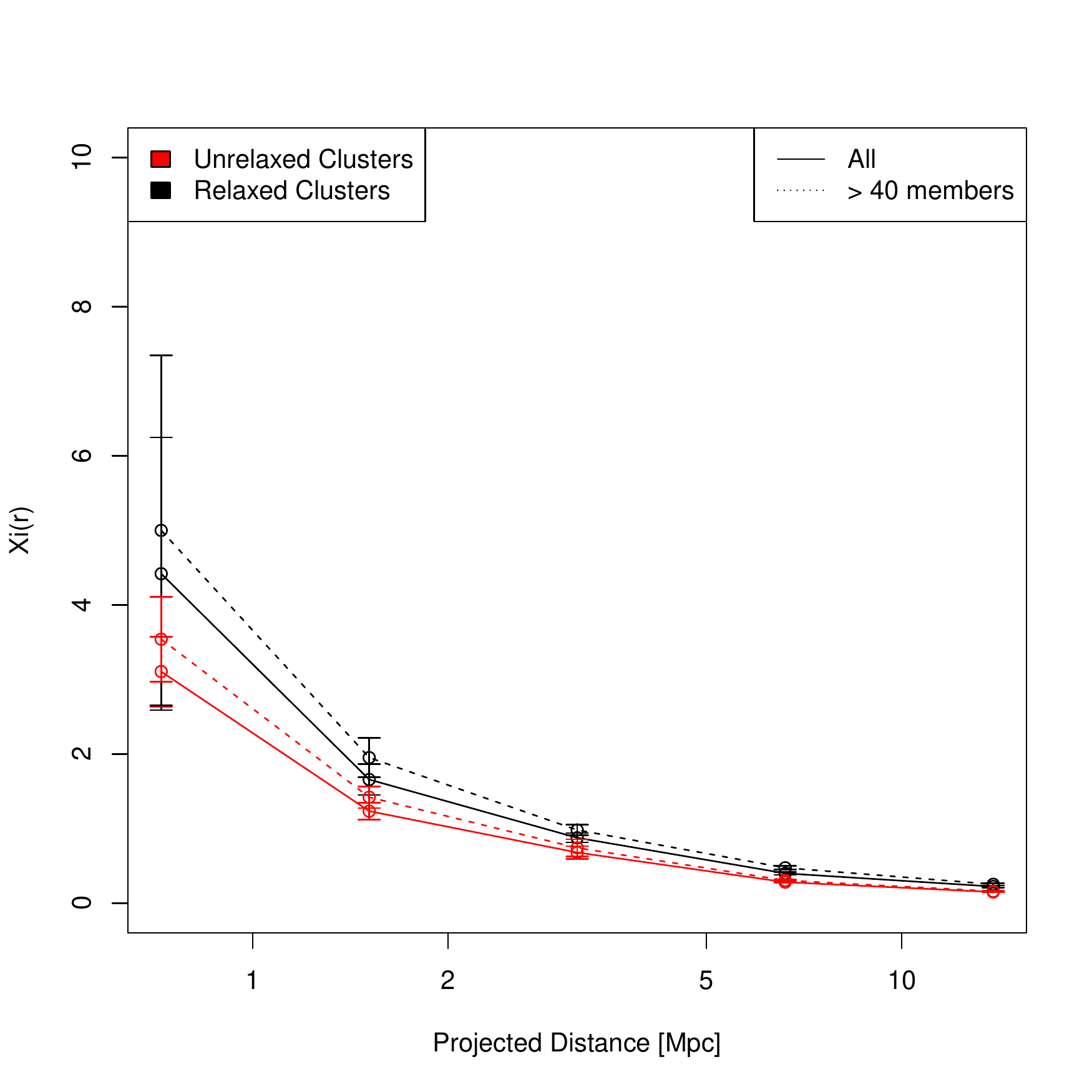}
\caption[projected Xi]
{Faraday rotation absolute value averaged as a function of projected distance to the galaxy cluster samples. 
Same colors and line reference as fig. \ref{fig:angular}.
}
\label{fig:Xi}
\end{figure}

Given the nature of the RM distribution \citep[which does not follow a Gaussian
shape for example]{taylor2009}, we analyzed our results with a different
statistical estimator to derive the relations between the cosmic structures and their magnetic component. 
In particular, we can define 

\begin{equation}
\frac{\xi(r)}{{\rm RM}} = < |RM| dr >  
\label{eqn:Xi} 
\end{equation}

which is averaging in projected distance bins of the absolute value of the
rotation measurement. 
In figure \ref{fig:Xi}, we show the result of this estimator. In this case, the
relaxed systems show a higher RM absolute value averaged towards the center in
comparison with the unrelaxed clusters. Meaning that even if they have a smaller
number of lines of sight RMs, we can
measure the high ones. Note that the errors shown in the plot are by bootstraping the
sample; therefore, the higher error bars for the relaxed clusters only reflect
the fact of small numbers of RMs.  This information tells us about the nature of
the distribution of RM in each radial bin, implying that the unrelaxed
distribution is skewed and has a larger dispersion given the nature of the
processes involved in the unrelaxed systems.

It is worth mentioning that, in order to build the catalog, \cite{taylor2009} used
data from the {\rm NVSS} that was not designed for polarization studies and,
given the beam size and sensitivity available, it suffered a depolarization
effect. Nevertheless, the difference between the relaxed and unrelaxed clusters
holds even using two different RM catalogs \citep{taylor2009,xu2014} with
different RM intrinsic errors and information from the sources.  It is also
worth to remark that although it is most likely that the relaxed sample has
clusters with a cool core component and thus a small-scale dynamo acting, our
studies focus at larger scales than the cluster center.

\section{Conclusions} \label{sec:conclusions}

Taking into account that the presence of magnetic fields in galaxy clusters is
a very well-known fact, and that the structure formation proceeds in a
hierarchical way giving rise to bigger structures through mergers, we studied
the relation between RM measurements and the dynamical status of the galaxy
clusters.  In order to do so, we used the galaxy clusters catalog constructed
by \cite{wen} and correlated the position of each cluster with RM sources.  We
found that the unrelaxed clusters have a significant higher RM standard
deviation $\sigma_{RM}$ than the relaxed ones.  This difference can be
interpreted as if the unrelaxed clusters have higher MF at
megaparsec scales with kiloparsecs coherence lengths.  These results can be
understood if we consider the small-scale dynamo caused by minor and major
mergers.

In order to see if this difference is really due to the dynamical state, we
divided the cluster sample into bright and faint.  We found that there
is no difference between the MFs of these two samples. This suggests that
the difference found in section \ref{sec:analysis} is really due to the
dynamical status of the clusters.

The research that we present links properties of the Gas with two samples that rely 
on a numerical algorithm that takes into account optical features of the 
system relaxation. In section \ref{sec:data}, we already mentioned the information 
available for the learning sample, 
which does not include the gas content in the analysis.
However, we do observe that the relaxations state independently correlates
with the gas and the MF evolution inside galaxy clusters.

We also found that the unrelaxed clusters are more likely to have a crossing RM
in one virial radius than a random sample. For the case of the relaxed clusters,
we found this less likely. Therefore, there should be a depolarization process
for the rotations measurement in those systems.

We were able to estimate some typical MF values for those systems.
This result is strongly dependent on some parameters, as well as the electron
density $N_e$, the coherence length of the ICM and the Faraday depth
\citep{Stasyszyn2010,Vacca2016}.  The selection process of the two samples is
only photometric and, therefore, there are still a lot of uncertainties on the
physical properties for each of the samples that should be addressed in the
future.  However, by using conservative values for those parameters, we infer
characteristic values of $B_\parallel = 0.96\pm 0.08~ \mu {\rm G}$  for the
unrelaxed clusters and $B_\parallel = 0.45 \pm 0.08 \mu {\rm G} $ for the relaxed 
ones. These amounts are lower limits to the absolute value for the 
MF in those samples and do not consider the fact that small-scale dynamo
can be efficient \citep{Vazza2018} but not measurable with this staking
methods.  Nevertheless, it is an evidence that unrelaxed systems could have a typical
mechanism to enhance magnetic fields with large characteristic coherence
lengths to be measured thought the Faraday rotation measurements. 

Next generation of instruments, new catalogs and deeper studies of the
dynamical samples will give us the detail to understand the nature of
cosmological magnetic fields, their origin and their role in the large-scale
structure.

\section*{Acknowledgements}

We want to thank Franco Vazza and Annalisa Bonafede for the useful comments and
discussions and Carolina Charalambous for extensive corrections.
The authors wish to thank the anonymous referees for valuable comments which improve the paper.
This research was partially supported by Consejo Nacional de Investigaciones 
Cient\'{\i}ficas y T\'ecnicas (CONICET, Argentina) 
and Secretar\'{\i}a de Ciencia y Tecnolog\'{\i}a de la Universidad Nacional 
de C\'ordoba (SeCyT-UNC, Argentina).

\bibliographystyle{mnras}
\bibliography{biblio}

\begin{thebibliography}{}
\makeatletter
\relax
\def\mn@urlcharsother{\let\do\@makeother \do\$\do\&\do\#\do\^\do\_\do\%\do\~}
\def\mn@doi{\begingroup\mn@urlcharsother \@ifnextchar [ {\mn@doi@}
  {\mn@doi@[]}}
\def\mn@doi@[#1]#2{\def\@tempa{#1}\ifx\@tempa\@empty \href
  {http://dx.doi.org/#2} {doi:#2}\else \href {http://dx.doi.org/#2} {#1}\fi
  \endgroup}
\def\mn@eprint#1#2{\mn@eprint@#1:#2::\@nil}
\def\mn@eprint@arXiv#1{\href {http://arxiv.org/abs/#1} {{\tt arXiv:#1}}}
\def\mn@eprint@dblp#1{\href {http://dblp.uni-trier.de/rec/bibtex/#1.xml}
  {dblp:#1}}
\def\mn@eprint@#1:#2:#3:#4\@nil{\def\@tempa {#1}\def\@tempb {#2}\def\@tempc
  {#3}\ifx \@tempc \@empty \let \@tempc \@tempb \let \@tempb \@tempa \fi \ifx
  \@tempb \@empty \def\@tempb {arXiv}\fi \@ifundefined
  {mn@eprint@\@tempb}{\@tempb:\@tempc}{\expandafter \expandafter \csname
  mn@eprint@\@tempb\endcsname \expandafter{\@tempc}}}

\bibitem[\protect\citeauthoryear{{Abdo} et~al.,}{{Abdo} et~al.}{2010}]{abdo}
{Abdo} A.~A.,  et~al., 2010, \mn@doi [\apj] {10.1088/0004-637X/715/1/429},
  \href {http://adsabs.harvard.edu/abs/2010ApJ...715..429A} {715, 429}

\bibitem[\protect\citeauthoryear{{Assef}, {Stern}, {Noirot}, {Jun}, {Cutri}  \&
  {Eisenhardt}}{{Assef} et~al.}{2018}]{wise}
{Assef} R.~J.,  {Stern} D.,  {Noirot} G.,  {Jun} H.~D.,  {Cutri} R.~M.,
  {Eisenhardt} P.~R.~M.,  2018, \mn@doi [The Astrophysical Journal Supplement
  Series] {10.3847/1538-4365/aaa00a}, \href
  {https://ui.adsabs.harvard.edu/#abs/2018ApJS..234...23A} {234, 23}

\bibitem[\protect\citeauthoryear{{Beck}}{{Beck}}{2009}]{Beck2009ASTRA}
{Beck} R.,  2009, \mn@doi [Astrophysics and Space Sciences Transactions]
  {10.5194/astra-5-43-2009}, \href
  {http://adsabs.harvard.edu/abs/2009ASTRA...5...43B} {5, 43}

\bibitem[\protect\citeauthoryear{{B{\"o}hringer} \& {Werner}}{{B{\"o}hringer}
  \& {Werner}}{2010}]{boehringer2010}
{B{\"o}hringer} H.,  {Werner} N.,  2010, \mn@doi [\aapr]
  {10.1007/s00159-009-0023-3}, \href
  {http://adsabs.harvard.edu/abs/2010A%26ARv..18..127B} {18, 127}

\bibitem[\protect\citeauthoryear{{B{\"o}hringer}, {Chon}  \&
  {Kronberg}}{{B{\"o}hringer} et~al.}{2016}]{boehringer2016}
{B{\"o}hringer} H.,  {Chon} G.,   {Kronberg} P.~P.,  2016, \mn@doi [\aap]
  {10.1051/0004-6361/201628873}, \href
  {http://adsabs.harvard.edu/abs/2016A%26A...596A..22B} {596, A22}

\bibitem[\protect\citeauthoryear{{Bonafede}, {Govoni}, {Feretti}, {Murgia},
  {Giovannini}  \& {Br{\"u}ggen}}{{Bonafede} et~al.}{2011}]{Bonafede2011}
{Bonafede} A.,  {Govoni} F.,  {Feretti} L.,  {Murgia} M.,  {Giovannini} G.,
  {Br{\"u}ggen} M.,  2011, \mn@doi [\aap] {10.1051/0004-6361/201016298}, \href
  {http://adsabs.harvard.edu/abs/2011A%26A...530A..24B} {530, A24}

\bibitem[\protect\citeauthoryear{{Bonafede} et~al.,}{{Bonafede}
  et~al.}{2015}]{2015aska.confE..95B}
{Bonafede} A.,  et~al., 2015, Advancing Astrophysics with the Square Kilometre
  Array (AASKA14), \href {http://adsabs.harvard.edu/abs/2015aska.confE..95B}
  {p.~95}

\bibitem[\protect\citeauthoryear{{Carilli} \& {Taylor}}{{Carilli} \&
  {Taylor}}{2002}]{carilli2002}
{Carilli} C.~L.,  {Taylor} G.~B.,  2002, \mn@doi [\araa]
  {10.1146/annurev.astro.40.060401.093852}, \href
  {http://adsabs.harvard.edu/abs/2002ARA%26A..40..319C} {40, 319}

\bibitem[\protect\citeauthoryear{{Clarke}, {Kronberg}  \&
  {B{\"o}hringer}}{{Clarke} et~al.}{2001}]{clarke2001}
{Clarke} T.~E.,  {Kronberg} P.~P.,   {B{\"o}hringer} H.,  2001, \mn@doi [\apjl]
  {10.1086/318896}, \href {http://adsabs.harvard.edu/abs/2001ApJ...547L.111C}
  {547, L111}

\bibitem[\protect\citeauthoryear{{Feretti}, {Giovannini}, {Govoni}  \&
  {Murgia}}{{Feretti} et~al.}{2012}]{feretti2012}
{Feretti} L.,  {Giovannini} G.,  {Govoni} F.,   {Murgia} M.,  2012, \mn@doi
  [\aapr] {10.1007/s00159-012-0054-z}, \href
  {http://adsabs.harvard.edu/abs/2012A%26ARv..20...54F} {20, 54}

\bibitem[\protect\citeauthoryear{{Horiuchi} et~al.,}{{Horiuchi}
  et~al.}{2004}]{horiuchi}
{Horiuchi} S.,  et~al., 2004, \mn@doi [\apj] {10.1086/424811}, \href
  {https://ui.adsabs.harvard.edu/#abs/2004ApJ...616..110H} {616, 110}

\bibitem[\protect\citeauthoryear{{Kale} et~al.,}{{Kale}
  et~al.}{2016}]{2016JApA...37...31K}
{Kale} R.,  et~al., 2016, \mn@doi [Journal of Astrophysics and Astronomy]
  {10.1007/s12036-016-9406-9}, \href
  {http://adsabs.harvard.edu/abs/2016JApA...37...31K} {37, 31}

\bibitem[\protect\citeauthoryear{{Kravtsov} \& {Borgani}}{{Kravtsov} \&
  {Borgani}}{2012}]{kravtsovBorgani2012}
{Kravtsov} A.~V.,  {Borgani} S.,  2012, \mn@doi [\araa]
  {10.1146/annurev-astro-081811-125502}, \href
  {http://adsabs.harvard.edu/abs/2012ARA%26A..50..353K} {50, 353}

\bibitem[\protect\citeauthoryear{{Kronberg}}{{Kronberg}}{2016}]{Kronberg2016}
{Kronberg} P.~P.,  2016, {Cosmic Magnetic Fields}.
UK: Cambridge University Press

\bibitem[\protect\citeauthoryear{{Marinacci} et~al.,}{{Marinacci}
  et~al.}{2017}]{2017arXiv170703396M}
{Marinacci} F.,  et~al., 2017, preprint, \href
  {http://adsabs.harvard.edu/abs/2017arXiv170703396M} {} (\mn@eprint {arXiv}
  {1707.03396})

\bibitem[\protect\citeauthoryear{{Pakmor}, {Marinacci}  \& {Springel}}{{Pakmor}
  et~al.}{2014}]{2014ApJ...783L..20P}
{Pakmor} R.,  {Marinacci} F.,   {Springel} V.,  2014, \mn@doi [\apjl]
  {10.1088/2041-8205/783/1/L20}, \href
  {http://adsabs.harvard.edu/abs/2014ApJ...783L..20P} {783, L20}

\bibitem[\protect\citeauthoryear{{Pakmor} et~al.,}{{Pakmor}
  et~al.}{2017}]{2017MNRAS.469.3185P}
{Pakmor} R.,  et~al., 2017, \mn@doi [\mnras] {10.1093/mnras/stx1074}, \href
  {http://adsabs.harvard.edu/abs/2017MNRAS.469.3185P} {469, 3185}

\bibitem[\protect\citeauthoryear{{Stasyszyn}, {Nuza}, {Dolag}, {Beck}  \&
  {Donnert}}{{Stasyszyn} et~al.}{2010}]{Stasyszyn2010}
{Stasyszyn} F.,  {Nuza} S.~E.,  {Dolag} K.,  {Beck} R.,   {Donnert} J.,  2010,
  \mn@doi [\mnras] {10.1111/j.1365-2966.2010.17166.x}, \href
  {http://adsabs.harvard.edu/abs/2010MNRAS.408..684S} {408, 684}

\bibitem[\protect\citeauthoryear{{Stasyszyn}, {Dolag}  \& {Beck}}{{Stasyszyn}
  et~al.}{2013}]{stasyszyn2013}
{Stasyszyn} F.~A.,  {Dolag} K.,   {Beck} A.~M.,  2013, \mn@doi [\mnras]
  {10.1093/mnras/sts018}, \href
  {http://adsabs.harvard.edu/abs/2013MNRAS.428...13S} {428, 13}

\bibitem[\protect\citeauthoryear{{Stefani} et~al.,}{{Stefani}
  et~al.}{2017}]{Experiment2017}
{Stefani} F.,  et~al., 2017, preprint, \href
  {http://adsabs.harvard.edu/abs/2017arXiv170508189S} {} (\mn@eprint {arXiv}
  {1705.08189})

\bibitem[\protect\citeauthoryear{Subramanian, Shukurov  \& Haugen}{Subramanian
  et~al.}{2006}]{Subramanian2006}
Subramanian K.,  Shukurov A.,   Haugen N. E.~L.,  2006, \mn@doi [Monthly
  Notices of the Royal Astronomical Society]
  {10.1111/j.1365-2966.2006.09918.x}, 366, 1437

\bibitem[\protect\citeauthoryear{{Taylor}, {Stil}  \& {Sunstrum}}{{Taylor}
  et~al.}{2009}]{taylor2009}
{Taylor} A.~R.,  {Stil} J.~M.,   {Sunstrum} C.,  2009, \mn@doi [\apj]
  {10.1088/0004-637X/702/2/1230}, \href
  {http://adsabs.harvard.edu/abs/2009ApJ...702.1230T} {702, 1230}

\bibitem[\protect\citeauthoryear{{Vacca} et~al.,}{{Vacca}
  et~al.}{2016}]{Vacca2016}
{Vacca} V.,  et~al., 2016, \mn@doi [\aap] {10.1051/0004-6361/201527291}, \href
  {http://adsabs.harvard.edu/abs/2016A%26A...591A..13V} {591, A13}

\bibitem[\protect\citeauthoryear{{Vazza}, {Ferrari}, {Bonafede}, {Br{\"u}ggen},
  {Gheller}, {Braun}  \& {Brown}}{{Vazza} et~al.}{2015}]{2015aska.confE..97V}
{Vazza} F.,  {Ferrari} C.,  {Bonafede} A.,  {Br{\"u}ggen} M.,  {Gheller} C.,
  {Braun} R.,   {Brown} S.,  2015, Advancing Astrophysics with the Square
  Kilometre Array (AASKA14), \href
  {http://adsabs.harvard.edu/abs/2015aska.confE..97V} {p.~97}

\bibitem[\protect\citeauthoryear{{Vazza}, {Brunetti}, {Br{\"u}ggen}  \&
  {Bonafede}}{{Vazza} et~al.}{2018}]{Vazza2018}
{Vazza} F.,  {Brunetti} G.,  {Br{\"u}ggen} M.,   {Bonafede} A.,  2018, \mn@doi
  [\mnras] {10.1093/mnras/stx2830}, \href
  {https://ui.adsabs.harvard.edu/\#abs/2018MNRAS.474.1672V} {474, 1672}

\bibitem[\protect\citeauthoryear{{V{\'e}ron-Cetty} \&
  {V{\'e}ron}}{{V{\'e}ron-Cetty} \& {V{\'e}ron}}{2010}]{Veron2010}
{V{\'e}ron-Cetty} M.-P.,  {V{\'e}ron} P.,  2010, \mn@doi [\aap]
  {10.1051/0004-6361/201014188}, \href
  {http://adsabs.harvard.edu/abs/2010A%26A...518A..10V} {518, A10}

\bibitem[\protect\citeauthoryear{{Wen} \& {Han}}{{Wen} \& {Han}}{2013}]{wen}
{Wen} Z.~L.,  {Han} J.~L.,  2013, \mn@doi [\mnras] {10.1093/mnras/stt1581},
  \href {http://adsabs.harvard.edu/abs/2013MNRAS.436..275W} {436, 275}

\bibitem[\protect\citeauthoryear{{Xu} \& {Han}}{{Xu} \& {Han}}{2014}]{xu2014}
{Xu} J.,  {Han} J.-L.,  2014, \mn@doi [Research in Astronomy and Astrophysics]
  {10.1088/1674-4527/14/8/005}, \href
  {http://adsabs.harvard.edu/abs/2014RAA....14..942X} {14, 942}

\makeatother
\end{thebibliography}

\end{document}